\documentstyle[psfig]{mn0}

\def\simg{\mathrel{\rlap{\raise 0.511ex \hbox{$>$}}{\lower 0.511ex \hbox{$\sim$}}}}
\def\siml{\mathrel{\rlap{\raise 0.511ex \hbox{$<$}}{\lower 0.511ex \hbox{$\sim$}}}}
\def\beq{\begin{equation}} \def\eeq{\end{equation}} 
\def\gt{\hspace*{-1mm}>\hspace*{-1mm}}
\def\lt{\hspace*{-1mm}<\hspace*{-1mm}}
\def\cgs{\,{\rm erg\; cm^{-2}\, s^{-1}}} \def\CGS{\frac{\rm erg}{\rm cm^2 s}}
\def\ks{{\rm ks}} \def\h1{\hspace*{-1mm}} \def\2h{\hspace*{-2mm}} 
\def\3h{\hspace*{-3mm}} \def\4h{\hspace*{-4mm}} \def\hh{\hspace*{-0.5mm}}
\def\z3{\left(\hh\frac{z+1}{3}\hh\right)}

\begin{document}

\title [GeV afterglows] {GeV emission from Gamma-Ray Burst afterglows}

\author[A. Panaitescu]{A. Panaitescu \\
       Space Science and Applications, MS D466, Los Alamos National Laboratory,
       Los Alamos, NM 87545, USA}

\maketitle

\begin{abstract}
\begin{small}
 We calculate the GeV afterglow emission expected from a few mechanisms related to 
GRBs and their afterglows. Given the brightness of the early X-ray afterglow emission 
measured by Swift/XRT, GLAST/LAT should detect the self-Compton emission from the 
forward-shock driven by the GRB ejecta into the circumburst medium. Novel features 
discovered by Swift in X-ray afterglows (plateaus and chromatic light-curve breaks) 
indicate the existence of a pair-enriched, relativistic outflow located behind the 
forward shock. Bulk and inverse-Compton upscattering of the prompt GRB emission by 
such outflows provide another source of GeV afterglow emission detectable by LAT. 
The large-angle burst emission and synchrotron forward-shock emission are, most likely, 
too dim at high photon energy to be observed by LAT. The spectral slope of the 
high-energy afterglow emission and its decay rate (if it can be measured) allow 
the identification of the mechanism producing the GeV transient emission following GRBs.
\end{small}
\end{abstract}


\section{Sources of GeV afterglow emission}
\label{intro}

 In Gamma-Ray Bursts (GRBs), there are two main sources of GeV afterglow photons. 
One is the {\sl prompt emission} itself, which lasts for $t_\gamma=3-300$ s (e.g. 
Sakamoto et al 2007, Willingale et al 2007), has a broken power-law spectrum $F_\nu$ 
peaking at $\epsilon_p \sim 100$ keV, above which $F_\nu \propto \nu^{-\beta_\gamma}$ 
with $\beta_\gamma \in (0.6,2.3)$ and a 25 keV -- 1 MeV fluence $\phi = 10^{-5\pm 1} 
\; {\rm erg\,cm^{-2}}$ (for long bursts) (e.g. Preece et al 2000). 
 The other one is the {\sl afterglow emission} which, at X-rays, exhibits a plateau
at 0.3--30 ks characterized by a 1 keV flux $(\nu F_\nu)_{1keV} (t=1\ks) = 10^{-11\pm 1} 
\cgs$, a power-law spectrum $F_\nu \propto \nu^{-\beta_x}$ with $\beta_x \in (0.5,2)$ 
(e.g. O'Brien et al 2006, Willingale et al 2007) and a power-law decay 
$F_\nu \propto t^{-\alpha_x}$ with $\alpha_x \in (0,1)$ (e.g. Nousek et al 2006).

 In addition to the above "direct" sources of high-energy (GeV) afterglow photons,
a fraction of the hard X-ray photons from burst and soft X-ray photons from afterglow 
can be upscattered in the GeV range either through inverse-Compton off relativistic 
electrons or bulk-scattering off a faster part of the outflow. 
 In this work, we evaluate the GeV afterglow fluxes expected from these direct and 
scattering mechanisms and compare them to the sensitivity of the Large Area Telescope 
(LAT) at 1 GeV for a 1 ks observation, $(\epsilon F_\epsilon)_{LAT} = 10^{-9} \cgs$, 
corresponding to 5 photons being collected. 

 Another mechanism directly related to those at work in GRBs and their afterglows,
which can produce {\sl afterglow} emission above 100 MeV is the inverse-Compton 
scattering of flaring internal-shock emission passing through the forward-shock 
(Fan \& Piran 2006; Wang, Li \& M\'esz\'aros 2006; Fan et al 2007, Galli \& Piro 2007). 
That scenario is not considered here. Other processes which rely on mechanisms that 
are not directly related to the production of the GRB and afterglow emissions (e.g. 
synchrotron emission from VHE protons, synchrotron radiation from pairs and muons 
produced in photo-hadronic processes)
and whose high-energy emission is highly uncertain, are also ignored.

\subsection{Large-angle GRB emission}

 Owing to relativistic beaming and geometrical curvature of spherical GRB source, 
during the burst, the observer receives emission mainly from a patch moving toward
the observer and extending an angle opening $\theta_{grb} = \Gamma_{grb}^{-1}$ as
seen from the center of fireball, where $\Gamma_{grb}$ is the Lorentz factor of the 
source. Although the burst emission mechanism ceases to operate after 10--100 s, 
the observer continues to receive emission from regions of the GRB source moving 
at angles $\theta \gt \theta_{grb}$ relative to the center--observer axis (Kumar 
\& Panaitescu 2000). This "large-angle" emission arrives at observer at time 
$t(\theta) = t_\gamma (\theta/\theta_{grb})^2$ and is relativistically boosted by
a factor ${\cal D} (\theta) = 2\Gamma_{grb}/(1+\Gamma_{grb}^2\theta^2)$, hence
${\cal D} (t) \simeq 2\Gamma_{grb}(t_\gamma/t)$ is progressively decreasing with
observer time. From here, it can be shown that the large-angle burst emission at
photon energy $\epsilon$ above the burst spectral peak energy $\epsilon_p$ is 
$F_\epsilon(t)= F_\epsilon^{(grb)} (t/t_\gamma)^{-2-\beta_\gamma}$, where 
$F_\epsilon^{(grb)}= F_p (\epsilon/\epsilon_p)^{-\beta_\gamma}$ is the 
flux at energy $\epsilon$ during the burst and $F_p$ is the burst average 
flux at the photon energy $\epsilon_p$, which can be approximated 
by $F_p \simeq \Phi /(\epsilon_p t_\gamma)$. Assuming that the power-law 
spectrum of the prompt emission extends up to above GeV photon energies, the 
large-angle burst emission at $\epsilon \gt \epsilon_p$ is given by
\beq
 (\epsilon F_\epsilon)^{(lae)} = \frac{\phi}{t_\gamma} 
    \left(\frac{\epsilon}{\epsilon_p}\right)^{1-\beta_\gamma}
    \left(\frac{t}{t_\gamma}\right)^{-2-\beta_\gamma}  \;.
\eeq
For the representative burst quantities given in \S\ref{intro}, the large-angle flux 
at photon energy $\epsilon = 1\, \epsilon_9$ GeV is
\beq
 (\epsilon F_\epsilon)^{(lae)} \h1 =  \h1
    10^{-6(\beta_\gamma+1)} \phi_{-5}\, t_{\gamma,1}^{\beta_\gamma+1}  
    \h1 \left( \h1 \frac{\epsilon_9}{\epsilon_{p,5}} \h1 \right)^{\h1 \beta_\gamma-1}
    \2h \left(\frac{t}{1\ks}\right)^{\h1 -2-\beta_\gamma} \2h \CGS
\label{lae}
\eeq
where the notation $x_n = x/(10^n\, {\rm cgs})$ was used, the photon energy being
measured in eV.

 Equation (\ref{lae}) shows that, at 1 ks after trigger, the GeV large-angle emission 
may be detected by LAT at 1 GeV for 
$(i)$ the hardest bursts ($\beta_\gamma=0.5$), for which 
  $(\epsilon F_\epsilon)_{1GeV} = 10^{-9} \phi_{-5} t_{\gamma,1}^{1.5} \cgs$, or for
$(ii)$ the harder ($\beta_\gamma=1$), longest, and brightest bursts ($t_\gamma \gt 100$ s, 
 $\phi \gt 10^{-4} {\rm erg\, cm^{-2}}$), 
 for which $(\epsilon F_\epsilon)_{1GeV} = 10^{-9} \phi_{-4} t_{\gamma,2}^2 \cgs$.
Bursts in either category are extremely rare (perhaps less than 1 percent of BATSE
GRBs have those properties), thus is seems very unlikely that the large-angle
burst emission could produce 1 GeV photons that LAT can detect at 1 ks from trigger. 
Owing to its very fast decay, there is a much better chance to detect the large-angle 
emission at earlier times (e.g. at 100 s).

\subsection{Primary afterglow emission}
\label{FS}

 If the soft X-ray (1--10 keV) power-law spectrum measured by the X-Ray Telescope 
(XRT) on Swift extends over more than 5 orders of magnitude in photon energy above
10 keV, then the observed X-ray flux 
\beq
 (\nu F_\nu)_{1keV} = 10^{-11\pm 1} (t/1\ks)^{-\alpha_x} \;\cgs
\label{Fx}
\eeq
implies a GeV afterglow flux $\epsilon F_\epsilon = (\nu F_\nu)_{1keV}
(\epsilon/1\,{\rm keV})^{1-\beta_x}$ which is
\beq
(\epsilon F_\epsilon)^{(aglow)} = 10^{-(5+6\beta_x)}  (\nu F_\nu)_{1keV,-11}
               \, \epsilon_9^{1-\beta_x} \; \cgs \;.
\label{fs}
\eeq
Therefore, LAT-detectable GeV emission arising from the same mechanism as that for 
the X-ray afterglow emission is obtained for $\beta_x \lt (4\pm 1)/6$. Such hard 
spectra are measured for about 10 percent of Swift afterglows. 

 If the X-ray afterglow is the synchrotron emission from the forward-shock driven
by the GRB ejecta into the circumburst medium (M\'esz\'aros \& Rees 1997), then
the assumption that the soft X-ray spectrum $F_\epsilon \propto \nu^{-\beta_x}$ extends
up to GeV energies is quite likely wrong, particularly for the harder afterglow
spectra ($\beta_x \lt 2/3$): 

 (1) If the soft X-ray band were above the "cooling" frequency ($\nu_c$) of the 
synchrotron spectrum, then the electrons accelerated by the forward-shock would 
have a power-law distribution with energy ($\gamma_e m_e c^2$) flatter than 
$dn_e/d\gamma_e \propto \gamma_e^{-4/3}$, which implies that $dn_e/d\gamma_e$ must have 
a cut-off at some electron energy for which the synchrotron characteristic frequency
is above X-ray. However, that the electrons radiating at 1 GeV must have an energy 
larger by a factor $10^3$ than those radiating at 1 keV, suggests that the synchrotron 
frequency cut-off associated with the maximum electron energy is below 1 GeV. 

 (2) If the soft X-ray range is below $\nu_c$, then the GeV flux depends on the exact 
location of $\nu_c$. For 10 GRB afterglows with a good coverage at radio, optical, 
and X-ray, through modelling their multiwavelength measurements, we have found shock 
microphysical parameters, shock kinetic energies, and ambient medium densities for 
which $\nu_c$ should be between optical and soft X-ray at 1 ks (Panaitescu 2005). 
This suggests that, if other afterglows have similar forward-shock parameters, 
their cooling frequency is never too much above the soft X-ray range. Taking into
account that the spectrum of the forward-shock synchrotron emission should be
$F_\epsilon \propto \nu^{-\beta_x-1/2}$ above $\nu_c$, it follows that, for $\nu_c$
around 100 keV, the GeV flux is a factor $\sim 100$ lower than that given in equation 
(\ref{fs}). Then, the GeV flux from the forward-shock synchrotron emission would 
be detected by LAT only for the hardest ($\beta_x = 0.5$) and brightest 
($(\nu F_\nu)_{1keV} = 10^{-10} \cgs$) afterglows observed by XRT.

 If the X-ray afterglow is the forward-shock synchrotron emission upscattered in 
a different part of the outflow (see \S\ref{bulk}), then the upscattered $\nu_c$
could be above 1 MeV and the likelihood of detecting GeV photons from the same
mechanism as that yielding the X-ray afterglow will be larger, although still 
restricted to a minority of afterglows with the hardest X-ray emission.

\subsection{Self-Compton scattering of afterglow emission}
\label{fsic}

 We consider now the upscattering of the X-ray synchrotron forward-shock emission 
by the same electrons, the model for GeV {\sl afterglow} emission that has received 
most attention so far (e.g. M\'esz\'aros \& Rees 1994, Panaitescu \& M\'esz\'aros 1997, 
Dermer, Chiang \& Mitman 2000, Zhang \& M\'esz\'aros 2001, Pe'er \& Waxman 2005, Fan
et al 2007). Galli \& Piro (2007) offer a detailed treatment of this emission 
and comparison with LAT's detection capability. In contrast to these works, we want 
to estimate the GeV inverse-Compton flux expected for the observed X-ray synchrotron 
forward-shock flux by making minimal use of the usual afterglow parameters. 
 
 At a photon energy above the peak frequency of the scattered emission 
($\epsilon_p^{(ic)}$), the flux of the inverse-Compton forward-shock emission is 
\beq
  (\epsilon F_\epsilon)^{(ic)} = (\epsilon F_\epsilon)_p^{(ic)}
            (\epsilon/\epsilon_p^{(ic)})^{1-\beta_x}
\eeq
where $F_{\epsilon,p}^{(ic)}$ is the peak flux of the inverse-Compton afterglow 
spectrum. The spectral quantities of the upscattered emission can be related to those 
of the synchrotron's: 
\beq
  \epsilon_p^{(ic)} = \gamma_e^2 \epsilon_p^{(sy)} \;, \quad
      F_{\epsilon,p}^{(ic)} = \tau_e F_{\epsilon,p}^{(sy)}
\eeq
where $\gamma_e$ is the Lorentz factor of the electrons which radiate at the peak 
frequency $\epsilon_p^{(sy)}$ and $\tau_e$ is the optical thickness of the 
forward-shock to electron scattering. The observed X-ray emission at 1 keV is
$(\nu F_\nu)_{1keV} = (\epsilon F_\epsilon)_p^{(sy)} 
(1\,{\rm keV}/\epsilon_p^{(sy)})^{1-\beta_x}$, thus  
\beq
  (\epsilon F_\epsilon)^{(ic)} = (\nu F_\nu)_{1keV}\, (\epsilon/1\,{\rm keV})^{1-\beta_x} 
            \gamma_e^{2\beta_x} \tau_e \;. 
\label{fic}
\eeq
 
 If the GRB progenitor is a Wolf-Rayet star, the mass swept-up by the forward-shock is 
$M_{fs} = (dM/dt)R/v = 6\times 10^{28} R_{16}$ g, assuming a typical mass-loss rate
$dM/dt = 10^{-5} M_\odot {\rm yr^{-1}}$ and wind velocity $v=10^3\, {\rm km\, s^{-1}}$).
Conservation of the forward-shock energy $E_{fs}$ during its interaction with the stellar 
wind, $E_{fs} = \Gamma_{fs}^2 M_{fs} c^2$, and the relation between the shocks radius
and the arrival time of photons emitted by the region moving at an angle $\Gamma_{fs}^{-1}$ 
relative to the direction toward the observer, $R_{fs} \simeq \Gamma_{fs}^2 ct$, lead to 
a Lorentz factor of the shocked circumburst medium
\beq
 \Gamma_{fs} = 42\, (E_{fs,53}/A_*)^{1/4}\, (t/1\ks)^{-1/4} 
\label{Gfs}
\eeq
and radius
\beq
 R_{fs} = 2.3 \times 10^{16} (E_{fs,53}/A_*)^{1/2}\, (t/1\ks)^{1/2}\;{\rm cm} 
\label{Rfs}
\eeq
for a burst at redshift $z=2$, where $A_* \equiv (dM/dt)/(10^{-5} M_\odot {\rm yr^{-1}})
\times (10^3\, {\rm km\, s^{-1}}/v)$ (for Galactic WR stars, the wind parameter $A_*$
varies from 0.3 to 2 -- Nugis \& Lamers 2000). 
Therefore, the forward-shock optical thickness to electron scattering is
\beq
  \tau_e = \frac{\sigma_e M_{fs}}{4\pi m_p R_{fs}^2} \simeq 
            10^{-5} E_{fs,53}^{-1/2} A_*^{3/2}\, (t/1\ks)^{-1/2} 
\label{taufs}
\eeq 
where $\sigma_e$ is the Thomson cross-section for electron scattering.

 The hardest X-ray spectra ($\beta_x \simeq 1/2$) measured by XRT indicate that, at 
least for some afterglow, the cooling frequency of the synchrotron emission is above 
X-ray. For simplicity, we assume that this generally valid, hence $\gamma_e$ of equation 
(\ref{fic})) is the typical Lorentz factor of the electrons accelerated by the forward-shock,  
parameterized by the fraction of the post-shock energy that the electrons acquire
(if they all had same Lorentz factor $\gamma_e$) : 
\beq
 \gamma_e\, m_e c^2 = \varepsilon_e\, \Gamma_{fs}\, m_p c^2 
\eeq
where $\Gamma_{fs} m_p c^2$ is its internal energy per proton. 
Then, equation (\ref{Gfs}) leads to
\beq
  \gamma_e = 3800\; (\varepsilon_e/0.05) (E_{fs,53}/A_*)^{1/4}\, (t/1\ks)^{-1/4} 
\label{gmel}
\eeq
where the fractional electron energy was normalized to the maximum value found by 
Panaitescu (2005) from modelling the afterglow emission. XRT observations show that 
the peak frequency of the synchrotron forward-shock spectrum ($\epsilon_p^{(sy)}$) 
is below 0.3 keV before 1 ks after trigger thus, for the electron Lorentz factor of 
equation (\ref{gmel}), the peak frequency of the upscattered emission ($\epsilon_p^{(ic)}$) 
at 1 ks should be below the GeV range. 

 Combining equations (\ref{fic}), (\ref{taufs}), and (\ref{gmel}), we obtain that,
for the X-ray flux typically measured by XRT (equation \ref{Fx}), the flux of the 
forward-shock self-Compton flux should be
\begin{displaymath} 
 (\epsilon F_\epsilon)^{(ic)} \leq  10^{-10 +1.2\beta_x}
     \left(\frac{\varepsilon_e}{0.05}\right)^{2\beta_x} E_{fs,53}^{\frac{\beta_x-1}{2}} 
            A_*^{\frac{3-\beta_x}{2}} 
\end{displaymath} 
\beq 
 \hspace*{15mm} \times (\nu F_\nu)_{1keV,-11} \epsilon_9^{1-\beta_x} 
      \left( \frac{t}{1\ks} \right)^{\h1-\frac{\beta_x+1}{2}}  \CGS 
\label{ic}
\eeq 
with a weak dependence on the forward-shock's kinetic energy.
Therefore, for the typical $\beta_x = 1$ measured by XRT, X-ray afterglows brighter
than average should be accompanied by inverse-Compton emission detectable to LAT.

\subsection{Bulk-scattering of GRB emission}
\label{bulk}

 Most of the X-ray afterglows monitored by Swift exhibit {\sl flares} and light-curve 
{\sl plateaus} at 0.1--10 ks after trigger. Most flares evolve on a timescale shorter 
than the time when they occur (Burrows et al 2007, Chincarini et al 2007), which 
indicates that they are not from the forward-shock. Instead, flares have been 
attributed (e.g. Zhang et al 2006) to the same mechanism (perhaps synchrotron and/or 
self-Compton emission from {\sl internal shocks} in a relativistic, unstable outflow 
-- Rees \& M\'esz\'aros 1994) that produces the highly-variable burst emission. 
Therefore, this scenario for X-ray flares require a long-lived central engine, 
expelling relativistic outflows at lab-frame times comparable to the observer time 
when the flares are seen.

 During the X-ray plateau, the 0.3--10 keV afterglow flux decays slower than 
expected for the forward-shock synchrotron emission, which prompted speculations that 
(at least one) of the basic assumptions is invalid. One plausible scenario is that 
energy is injected into the forward-shock by means of some late ejecta which catch-up 
with the forward shock as the latter undergoes deceleration by sweeping-up the ambient 
medium (Nousek et al 2006, Panaitescu et al 2006a, Zhang et al 2006). In that case, 
the plateau end marks a change in (e.g. cessation of) the power injected in the 
forward-shock. However, this interpretation encounters the following problem: a change 
in the forward-shock dynamics should be manifested in the afterglow light-curve at 
all frequencies, yet about half of the plateau ends which have been monitored in the 
optical do not occur in the optical as well, i.e. are {\sl chromatic} (Watson et al 
2006, Panaitescu et al 2006b). An alternative explanation for such chromatic X-ray
light-curve breaks -- the passage of a spectral break through the X-ray -- is ruled
out by that the hardness of the X-ray continuum does not evolve across the light-curve 
break (Willingale et al 2007, Liang et al 2007). 

 Thus, the existence of X-ray flares and plateaus suggest a sustained ejection of 
relativistic material, but the chromatic X-ray breaks cannot be explained by the
injection of energy into the forward-shock produced by the late outflow catching-up 
with the shock. Instead, those chromatic X-ray breaks suggest that there is a 
contribution to the afterglow flux that does not arise from the forward shock.

 As shown by Panaitescu (2007), bulk scattering of the forward-shock emission off 
the extended outflow required by X-ray flares can account for the observed decoupling 
of the optical and X-ray emissions because the scattered emission is likely to overshine 
that coming directly from the forward-shock only at higher (X-ray) but not at lower 
(optical) photon energies. 
For this to happen, the scattering outflow, inner to the forward-shock, must 
(1) move at a higher Lorentz factor (which is a natural consequence of the 
    forward-shock deceleration caused by its interaction with the circumburst medium) 
    and, quite likely, 
(2) should be enriched with leptons above what is expected for normal, baryonic ejecta
    (which suggests that the scattering outflow is not accelerated from within the
    collapsing star but, instead, it results from the dissipation of magnetic fields
    at large distances from the progenitor -- the electromagnetic model of Lyutikov 2006a).
The first condition above implies that the brightness of the scattered emission received 
by observer at time $t$ depends on the Lorentz factor and mass-flux (i.e. optical thickness) 
in the scattering outflow at a distance $ct$ behind the forward-shock, consequently the 
observer-frame duration of the X-ray light-curve plateau is equal to the lab-frame
light travel-time between the inner edge of the scattering outflow and the GRB source.

 In addition to X-ray plateaus and chromatic breaks, the scattering model can also explain 
the short duration of X-ray flares (Shen et al 2007 have investigated the brightness
of X-ray flares resulting from scattering the burst emission off geometrically thin, 
relativistic outflows) and the observed diversity of post-plateau decays, some of which
display very fast decays ($t^{-3}$ to $t^{-9}$). The former arises from that the reflection 
of the photons emitted by the forward-shock (moving at $\Gamma_{fs}$) off a surface 
moving at a higher Lorentz factor ($\Gamma_{sc}$) reduces observer-frame timescales by 
a factor $\Gamma_{fs}^2/\Gamma_{sc}^2$. The latter feature is naturally accommodated 
because the decay of the scattered emission depends on the radial distribution of 
mass-flux and Lorentz factor in the scattering outflow. 
 
 A pair-rich outflow located behind the forward-shock provides two ways to produce GeV 
afterglow photons. One is the scattering of X-ray forward-shock photons to higher 
energies. If the X-ray-to-GeV photons arise from this mechanism then the fluxes
at low and high photon energies are related through equation (\ref{fs}). As noted in 
\S\ref{FS}, in this model, the upscattered cooling frequency could be above 1 MeV, 
thus the scattered forward-shock emission of the brightest and hardest X-ray afterglows 
may be detected by LAT. 

 GeV photons are also produced through scattering of the prompt, GRB emission.
The dependence of the scattered GRB emission on the various burst and scattering
outflow parameters can be obtained as follows. For seed photons moving radially inward 
and scattered photons moving radially outward along the observer--center of explosion,
the Doppler factor for the scatterer--GRB source relative motion is ${\cal D} = 
\Gamma_{sc}/\Gamma_{grb}$. Taking into account that 
$(i)$ the observed scattered and seed photon energies are a larger by a factor 
  $\Gamma_{sc}$ and $\Gamma_{grb}$, respectively, than in the corresponding source 
  frame,  
$(ii)$ bulk-scattering increases the energy of the seed photon by a factor ${\cal D}$,
and 
$(iii)$ if the scattering electrons are hot (of random Lorentz factor $\gamma_e$), then
 inverse-Compton scattering boosts the energy of the primary photon by a factor 
 $\gamma_e^2$ (in the frame of the scattering outflow), 
it follows that the peak of the scattered emission is at a photon energy
\beq
 \epsilon_p^{(sc)} = (\Gamma_{sc} \gamma_e / \Gamma_{grb})^2  \epsilon_p 
\label{scnup}
\eeq
where $\epsilon_p$ is the peak photon energy of the burst spectrum. 
If the scatterer were infinitesimally thin (a surface), then the spectral peak flux of 
the scattered emission would be 
\beq
 F_p^{(sc)} = (1-e^{-\tau_e}) (\Gamma_{sc}/\Gamma_{grb})^2 F_p
\label{fsc}
\eeq
where $F_p$ is the burst flux at the peak of its spectrum. 
The first factor represents the fraction of incident photons that are upscattered, 
the second is the product of a factor $\Gamma_{sc}/\Gamma_{grb}$ accounting for the 
effect of time contraction on the received peak-fluxes and a factor ${\cal D}$ 
accounting for the effect of scattering due to the relative motion of the scatterer 
and GRB source.
We note that the motion of the two sources toward the observer does enhance the
comoving specific intensities by factors $\Gamma_{sc}^3$ and $\Gamma_{grb}^2$
(one factor for time contraction, two powers for angular beaming), however the
received fluxes "lose" the angular-beaming factor $\Gamma^2$  because this effect 
also reduces the fractional area of the spherical source whose emission is beamed 
toward the observer by a factor $\Gamma^2$.
The relative motion of the scatterer and GRB source enhances the intensity of the
incident flux by a factor ${\cal D}^3$; however, in the scatterer's frame, the 
highly collimated incident flux, arriving from within ${\cal D}^{-2}$ 
around the radial direction of motion (owing to relativistic beaming) is nearly 
isotropically redistributed by electron scattering, hence the net increase in 
intensity produced by scattering is just a factor ${\cal D}$.

 The GRB emission scattered by a surface is spread (in the observer frame) over
a time $\delta t_{sc} = (\Gamma_{grb}/\Gamma_{sc})^2 t_\gamma$ (this is both the
angular spread in the photon arrival time and the time to sweep-up the photons
released by the GRB source). For a radially-extended scattering outflow, the scattered 
GRB emission is further spread over an observer-frame equal to the light travel-time
from the inner to the outer edges of the scattering outflow. As the X-ray plateau
results from upscattering by the same outflow (but of the forward-shock emission),
the geometrical thickness of the scattering outflow is approximately equal to the 
duration $t_X$ of the X-ray plateau. 
Therefore, for $\Gamma_{sc} \simg \Gamma_{grb}$, a burst duration $t_\gamma \siml 300$ 
s, and an X-ray plateau lasting for $t_X \sim 10$ ks, the photon arrival-time spread 
arising from the radial extent of the scattering outflow is dominant and the received 
flux is lower than that given in equation (\ref{fsc}) by a factor $\delta t_{sc}/t_X$. 

 The scattering outflow lags by $t_X \lt 10$ ks behind the GRB source, whose radius is 
$R_\gamma/c = \Gamma_{grb}^2 t_\gamma = 100\, \Gamma_{grb,2}^2 t_{\gamma,1}$ ks
\footnote{ A GRB source radius larger than $10^{15}$ cm is obtained by Lyutikov (2006b)
 from the timing of the GRB tail, assuming $\Gamma_{grb} \simg 100$. A similarly
 large burst radius is obtained by Kumar et al (2007) for a set of 10 GRBs from the 
 the GRB tail epoch and from that the forward-shock is already decelerating when the 
 X-ray afterglow emission emerges. 
 Then, the $10^{-3}-1$ s variability timescale typically observed can be accommodated 
 if the GRB emission arises from hot-spots whose angular extent is (much) smaller than 
 the visible area of opening $\Gamma_{grb}^{-1}$. 
 Alternatively, if the burst emission arises from the entire $\Gamma_{grb}^{-1}$ visible 
 area, the burst variability timescale and the assumed GRB radius require a $\Gamma_{grb}$
 larger by a factor 3--100 than considered here and a correspondingly larger $\Gamma_{sc}$.
 } .
Thus, the radius of the scattering outflow can be approximated by that of the GRB 
source, hence the evolution of the scatterer's optical thickness ($\tau_e \propto
R_\gamma^{-2}$) is
\beq
 \tau_e \h1 = \h1 \frac{t_0^2}{t^2}, \; t_0 \h1 \equiv \h1 \left( \frac{\sigma_e E_{sc}}
    { 4\pi\,\Gamma_{sc} \gamma_e \Gamma_{grb}^4 m_e c^4} \right)^{\h1 1/2} \4h
   =  \frac{2.7\,{\rm s}}{\Gamma_{grb,2}^2} \h1 \left( \frac{E_{sc,53}}{(\Gamma_{sc} 
      \gamma_e)_4} \right)^{\h1 1/2} 
\label{taut}
\eeq
for a pair-dominated outflow of kinetic energy $E_{sc}$. The product $\Gamma_{sc} 
\gamma_e$ was scaled to $10^4$, for which the scattered spectrum peaks at 1 GeV 
(equation \ref{scnup}, with $\Gamma_{grb} = 100$ and $\epsilon_p \simeq 100$ keV). 
By integrating $1-e^{-\tau_e}$ over the evolution given in equation (\ref{taut}), 
it can be shown that the average fraction of scattered photons is $2t_0/t_\gamma$, 
for $t_0 \lt t_\gamma$.

 Putting together all these factors, the peak flux of the scattered emission 
(equation \ref{fsc}) satisfies
\beq
 F_p^{(sc)} \propto \frac{t_0}{t_\gamma} \left( \frac{\Gamma_{sc}}{\Gamma_{grb}} 
               \right)^2 \frac{\delta t_{sc}}{t_X} F_p = \frac{t_0}{t_X} F_p
\label{scfp}
\eeq
The spectrum of the scattered emission above its peak $\epsilon_p^{(sc)}$ has the 
same slope as the burst spectrum above $\epsilon_p$, thus $F_\epsilon^{(sc)} \propto
F_p^{(sc)} (\epsilon/\epsilon_p^{(sc)})^{-\beta_\gamma}$. Then equations (\ref{scnup}) 
and (\ref{scfp}) for the spectral characteristics of the scattered emission spectrum 
lead to 
\beq
 (\epsilon F_\epsilon)^{(sc)} \propto \frac{\phi}{t_\gamma} \frac{E_{sc}^{1/2}}{t_X} 
    \frac{(\Gamma_{sc}\gamma_e)^{2\beta_\gamma-1/2}} {\Gamma_{grb}^{2\beta_\gamma+2}} 
    \left(\frac{\epsilon_p}{\epsilon} \right)^{\beta_\gamma-1} 
\label{sc}
\eeq
where $\phi$ is the GRB fluence and $\epsilon_p F_p \simeq \phi/t_\gamma$ was used.

\begin{figure*}
\centerline{\psfig{figure=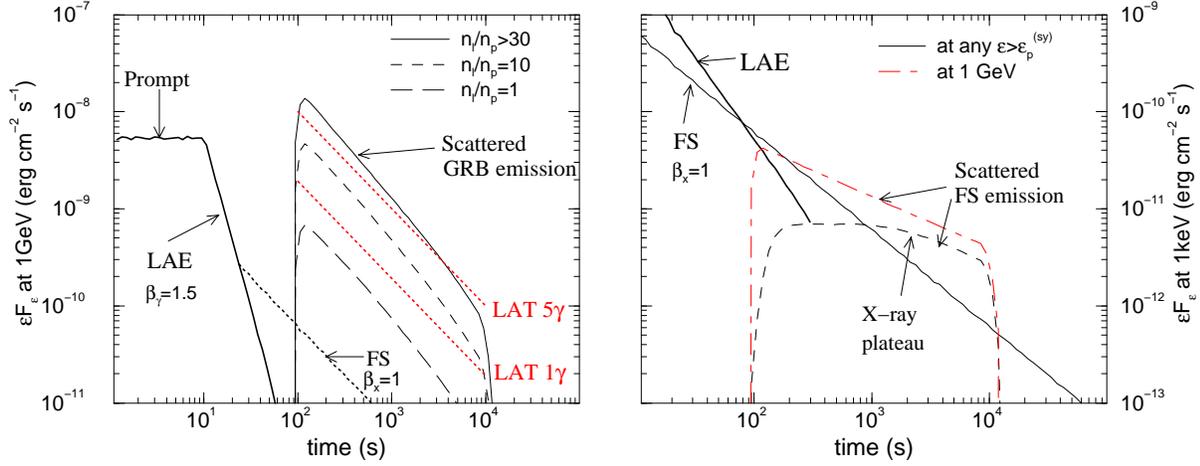,width=16cm}}
\caption{ 
  {\bf Right panel}: 
   X-ray plateau and 1 GeV emission from scattering the {\sl forward-shock} (FS) 
    synchrotron emission by a delayed, lepton-rich ($n_{l(epton)} \gg n_{p(roton)}$), 
    relativistic outflow. Also shown are the X-ray direct FS emissions (for spectral
    slope $\beta_x=1$) and the delayed {\sl large-angle emission} (LAE) released during 
    the burst but arriving later at observer.
   The forward-shock energy is $10^{53}$ ergs (comparable to the GRB output) and the 
    circumburst medium has the radial density distribution expected for a Wolf-Rayet wind.
   The scattering outflow properties (see below) were chosen so that the scattered flux 
    overshines that of the forward shock and yields a flux of  $10^{-11} \cgs$ at 1 keV
    with a spectral slope $\beta_x=1$ (average values measured by XRT).  
   In this case, neither the direct synchrotron FS nor the scattered emission are 
    detectable by LAT. 
   {\bf Left panel}: 
    GeV emission resulting from scattering the $\sim 100$ keV {\sl burst} emission 
     can be detected by LAT if the scattering outflow is pair-enriched ($n_l \gt 10\, n_p$),
     as required to produce an X-ray plateau. Red dotted lines labelled "LAT" indicate 
     the 1 GeV flux corresponding to LAT detecting 1 and 5 photons for an integration 
     time equal to the time given on abscissa.
    The burst spectral peak flux was set to 2 mJy which, for a 10 s burst with a 
     $F_\epsilon \propto \epsilon^{1/3}$ below its 100 keV peak and $F_\epsilon 
     \propto \epsilon^{-1.5}$ above it, corresponds to a 25--1000 keV fluence of 
     $10^{-5}\, {\rm erg\;cm^{-2}}$ (average parameters for BATSE long bursts). 
     The Lorentz factor of the GRB source (ahead of the scattering outflow) is 100.
   {\bf Both panels}: 
    The scattering outflow has a kinetic energy of $10^{53}$ ergs, a scattering 
     parameter $\Gamma \gamma_e = 10^4$ and is radially extended, being initially 
     $10^2-10^4$ light-seconds behind the forward shock. This gap introduces a delay 
     between the arrival times of the direct and scattered burst emissions. 
     The radial distribution of scattering outflow's Lorentz factor and mass were 
     assumed to be uniform. 
    }
\label{scatter}
\end{figure*}

 In the derivation of equation (\ref{sc}), we have considered only scattered photons
that move along the observer--center direction, for which ${\cal D} = \Gamma_{sc}/
\Gamma_{grb}$. For incident photons moving along other directions and being scattered 
toward the observer, the relativistic boost associated with the GRB source--scatterer
relative motion is much smaller.
In the numerical integration of the scattered emission, we take into account that 
dependence of relativistic beaming of the incoming GRB emission on the angle at 
which the photon moves (relative to the radial direction), as well as other factors
(e.g. the decrease of the scatterer's optical thickness as it expands, the effect 
of the source's geometrical curvature on the photon arrival time and energy).
The numerical calculation of the scattered GRB emission confirms the dependence of
the scattered flux on the model parameters given in equation (\ref{sc}), sets its
normalization and temporal evolution (Figure \ref{scatter}):
\beq
 (\epsilon F_\epsilon)^{(sc)} \h1 \hh = \h1 10^{-9} \frac{\phi_{-5}}{t_{\gamma,1}}  
      \frac{E_{sc,53}^{1/2}}{t_{X,4}} 
  \frac{(\Gamma_{\hh sc}\hh\gamma_e)_4^{2\beta_\gamma\h1-0.5}}
        {\Gamma_{grb,2}^{2\beta_\gamma+2}} 
   \h1 \left(\h1 \frac{\epsilon_9}{\epsilon_{p,5}} \h1\right)^{\h1 \hh 1 \hh -\beta_\gamma}
    \2h \left(\h1 \frac{t}{1\ks} \h1\right)^{\h1- \hh 1} \h1 \CGS .
\label{grbsc}
\eeq
 The above decay of the scattered emission is obtained for a uniform scattering outflow,
in which the ejecta mass $dM/dt$ and Lorentz factor $\Gamma_{sc}$ are constant. 
Various decays can be obtained if these quantities vary with the depth $ct$ in the 
scattering outflow, e.g. for $dM/dt \propto t^m$ and $\Gamma_{sc} \propto t^g$, the
scattered flux above $\epsilon_p^{(sc)}$ evolves as $F_\epsilon^{(sc)} \propto
t^{m+(2\beta_\gamma-0.5)g-1}$.

 Equation (\ref{grbsc}) indicates that LAT will detect the scattered GRB emission for 
(1) brighter bursts ($\phi/t_\gamma \gt 10^{-6} \cgs$) and/or (2) afterglows with 
shorter plateaus ($t_X \lt 10$ ks).

\section{Pair-formation opacity for GeV photons}

 For a test photon of energy $\epsilon$ emitted at radius $R$ by a source moving at 
Lorentz factor $\Gamma$, most of the optical thickness to pair-formation on photons
{\sl from same source} arises as the test photon travels a distance of order $R$ 
(because of the dilution of photons -- $n_\gamma \propto R^{-2}$). The relativistic 
motion of the source collimates most photons within an angle $\theta = \Gamma^{-1}$ 
around the radial direction. For this reason, during a small radial displacement $dr$, 
the test photon crosses a distance $dr/(2\Gamma^2)$ of the photon front (defined by 
the location of photons moving at angle $\Gamma^{-1}$ relative to the test photon), 
thus the corresponding infinitesimal optical thickness to pair-formation  
$d\tau_{\gamma\gamma} = (dr/2\Gamma^2) \sigma_{\gamma\gamma} n(\epsilon)$, where 
$n(\epsilon)$ is the lab-frame density of the target photons with which the test 
photon can form pairs. Integrating from $R$ to $2R$ over the dilution of target 
photons and using the upper limit of $\simeq (1/4) \sigma_e$ for the cross-section 
$\sigma_{\gamma\gamma}$ for pair-formation, we obtain
\beq
 \tau_{\gamma\gamma}(\epsilon) \siml \frac{\sigma_e N[>\epsilon_t(\epsilon)]} {32\pi\,R^2}
\label{taugg}
\eeq
where $N(>\epsilon_t)$ is the number of photons above the threshold energy 
$\epsilon_t$ for pair-formation on a test photon of energy $\epsilon$, contained 
in a slab of thickness $R/(2\Gamma^2)$ ahead of the shell at radius $R$ (i.e. the 
number of photons emitted from observer-frame time $t(R)$ to $2t(R)$).

 We estimate now the number $N(>\epsilon_t)$ of target photons emitted by a source 
of flux $(\epsilon F_\epsilon)_{1GeV} = 10^{-9} (\epsilon F_\epsilon)_{-9}$ $\cgs$ 
(normalization chosen for 5 photons collected by LAT in a 1 ks observation) during
$t-2t$, for an observer-frame test photon of energy $\epsilon = 1\, \epsilon_9$ GeV.

 For a high-energy flux with the above properties, the number of photons with lab-frame
energy above $(z+1) \epsilon$ emitted from observer time $t$ to $2t$ is
\beq
 N[>(z+1)\epsilon] \h1 = \h1 \frac{4\pi d_l^2(z) \epsilon F_\epsilon}{(z+1)\epsilon}
        \frac{t}{z+1}
        \h1 = \h1 10^{54.3} \h1 \z3^{\h1 2} \frac{(\epsilon F_\epsilon)_{-9}}{\epsilon_9} 
        \frac{t}{1\ks}
\label{N0}
\eeq
where $d_l \simeq 5\times 10^{27} (z+1)^2$cm is the luminosity distance; this approximation
is within 25 percent of the correct value for $z \in (0.5,5)$ .
 The condition for pair-formation $\epsilon_{target} \epsilon_{test} (1-\cos\theta) \geq 
2 (m_ec^2)^2$ and that most target photons move at angle $\theta \siml \Gamma^{-1}$ 
relative to the test photon, imply that the lab-frame minimum energy $\epsilon_t$ of a 
target photon with which the test photon of observer-frame energy $\epsilon$ can form a 
pair is
\beq
 \epsilon_t (\epsilon) = 350\, [(z+1)/3]^{-1} \Gamma^2 \epsilon_9^{-1} \; {\rm eV} \;.
\label{et}
\eeq
 For a $F_\epsilon \propto \epsilon^{-\beta}$ spectrum, the number of photons with 
energy above $\epsilon_t$ is
\begin{displaymath}
 N[>\epsilon_t(\epsilon)] = [\epsilon_t/(z+1)\epsilon]^{-\beta} N[>(z+1)\epsilon] 
\end{displaymath}
\beq
    \hspace*{5mm} = 10^{54.3+7\beta} [(z+1)/3]^{2\beta+2} \Gamma^{-2\beta}  
         (\epsilon F_\epsilon)_{-9} (t/1\ks) \,\epsilon_9^{2\beta-1} .
\label{N}
\eeq

 To complete the calculation of the optical thickness to pair-formation, the radius 
and Lorentz factor of the GeV source must be specified. We restrict attention to the
two mechanisms which are most likely to yield a LAT-detectable emission at 1 GeV:
inverse-Compton scatterings in the forward shock (\S\ref{fsic}) and bulk-scattering 
of the burst emission in a delayed outflow (\S\ref{bulk}).

\subsection{Self-Compton forward-shock emission}

 For the forward-shock Lorentz factor given in equation (\ref{Gfs}), the lab-frame
threshold energy of the target photon (equation \ref{et}) with which a test photon 
can annihilate is 
\beq
 \epsilon_t=0.6\,[(z+1)/3]^{-1/2} (E_{fs,53}/A_*)^{1/2} (t/1\ks)^{-1/2} \epsilon_9^{-1} 
  \;{\rm MeV} 
\label{etfs}
\eeq
which is around 200 keV in the observer frame. The peak energy of the inverse-Compton 
spectrum ($\epsilon_p^{(ic)}$) is very likely below 1 GeV but may be above the 
threshold energy $\epsilon_t$. For $\epsilon_p^{(ic)} \gt \epsilon_t$, the number of 
photons above the threshold $\epsilon_t$ increases with decreasing spectral peak 
energy $\epsilon_p^{(ic)}$, thus an upper limit on the optical thickness to 
pair-formation is obtained by assuming that $\epsilon_p^{(ic)} \lt \epsilon_t$. 
Then, the number of target photons is obtained by substituting the forward-shock 
Lorentz factor of equation (\ref{Gfs}) in equation (\ref{N}): 
$N(>\epsilon_t) \lt 10^{54.3+3.7\beta_x} [(z+1)/3]^{1.5\beta_x+2} 
(\epsilon F_\epsilon)_{-9} (A_*/E_{fs,53})^{\frac{1}{2}\beta_x} \\ 
(t/1\ks)^{\frac{1}{2}(\beta_x+1)}$  $\epsilon_9^{2\beta_x-1}$. 
 Together with the forward-shock radius given by equation (\ref{Rfs}), the optical
thickness to pair-formation of equation (\ref{taugg}) is
\begin{displaymath}
 \tau_{ic,ic} < 10^{3.7\beta_x\hh-4.6} \hh \z3^{\h1 \frac{3}{2}\beta_x \hh+3} 
       \h1 (\epsilon F_\epsilon)_{-9} 
      \left(\frac{A_*}{E_{fs,53}}\right)^{\h1 \frac{1}{2}\beta_x+1}  
\end{displaymath}
\beq
      \hspace*{10mm} \times (t/1\ks)^{\frac{1}{2}(\beta_x-1)} \epsilon_9^{2\beta_x-1} .
\label{tauic}
\eeq
Thus, pair-opacity for the inverse-Compton forward-shock emission at 1 GeV could be 
important in soft X-ray afterglows ($\beta_x \gt 5/4$) if the peak energy of the 
upscattered emission spectrum is at only 200 keV. We note that, if the forward-shock
microphysical parameters inferred by modelling the broadband emission of 10 afterglows 
(Panaitescu 2005) are representative, then we expect the peak of the inverse-Compton
spectrum to be above 1 MeV at 1 ks. For a tenfold increase of the upscattered peak 
energy, the optical thickness of equation (\ref{tauic}) decreases by a factor 
$10^{\beta_x}$, thus it seems quite likely that, even for the hardest observed
X-ray afterglows ($\beta_x \simeq 1.5$), the self-Compton scattering emission is 
optically thin to pair-formation at 1 GeV.

 The test photon may also form pair on a forward-shock synchrotron photon. If the
X-ray plateau emission is synchrotron from forward-shock, then the number of these 
photons can be calculated by using the typical X-ray plateau flux (equation \ref{Fx}) 
in equation (\ref{N0}). Then, the number of target photons above the threshold energy 
given in equation (\ref{etfs}) is $N_{sy}(\gt\epsilon_t) = 10^{58.3-2.3\beta_x} 
[(z+1)/3]^{1.5\beta_x+2}$ $(\nu F_\nu)_{1keV,-11} E_{fs,53}^{-\beta_x/2} 
(t/1\ks)^{\frac{1}{2}(1+\beta_x)} \epsilon_9^{\beta_x}$, from where the optical 
thickness to pair-formation on forward-shock synchrotron photons is
\begin{displaymath}
 \tau_{ic,sy} < 10^{\h1-2.3\beta_x-0.6} \hh \z3^{\h1\frac{3}{2}\beta_x\hh+3} 
        \left(\frac{A_*}{E_{fs,53}}\right)^{\h1 \frac{1}{2}\beta_x+1}
\end{displaymath}
\beq
    \hspace*{10mm} \times  (\nu F_\nu)_{1keV,-11} (t/1\ks)^{\frac{1}{2}(\beta_x-1)} 
        \epsilon_9^{\beta_x} \;.
\label{tausy}
\eeq
Equations (\ref{tauic}) and (\ref{tausy}), with the GeV flux of equation (\ref{ic}),
show that pair-formation on forward-shock synchrotron photons could be more important 
than on inverse-Compton photons for the harder X-ray afterglows with $\beta_x \lt 0.7$,
provided that the forward-shock synchrotron spectrum extends above the threshold
energy $\epsilon_t$ of equation (\ref{etfs}) and the self-Compton spectrum peaks
below the same $\epsilon_t$. 

 The Klein-Nishina effect reduces the scattered flux at photon energies approaching 
the observer frame energy of the scattering electron, i.e. below $\Gamma_{fs} \gamma_e 
m_e c^2/(z+1)$. Equations (\ref{Gfs}) and (\ref{gmel}) yield a KN cut-off energy 
$\epsilon_* = 30$ $ [(z+1)/3]^{-1/2} (\varepsilon_e/0.05) (E_{fs,53}/A_* t_3)^{1/2}$GeV. 
Because the electrons accelerated at the forward-shock have a power-law distribution
with energy above that given in equation (\ref{gmel}), the actual cut-off of the
inverse-Compton spectrum may be well above $\epsilon_*$.

\subsection{Scattered burst emission} 

 For a scattering outflow that is more relativistic than the GRB source ($\Gamma_{sc} 
\simg 100\, \Gamma_{grb,2}$), equation (\ref{et}) shows that the threshold energy
for the target photon is $\epsilon_t = 350\, \Gamma_{sc,3}^2 \epsilon_9$ MeV. 
Substituting the GRB source radius 
\beq
  R_{grb} = \Gamma_{grb}^2 \frac{ct_\gamma}{z+1} = 
      10^{15}\; \Gamma_{grb,2}^2 \frac{t_{\gamma,1}}{(z+1)/3} \; {\rm cm}
\label{Rgrb}
\eeq
in equation (\ref{taugg}) and using equation (\ref{N}) with $\Gamma = \Gamma_{grb}$, 
the optical thickness to pair-formation for the scattered GRB emission is
\beq
 \tau_{sc,sc} \siml 10^{\beta_\gamma-1.9} 
        \h1 \z3^{\h1 2\beta_\gamma \hh+4} \h1 \frac{(\epsilon F_\epsilon)_{-9}} 
        {\Gamma_{sc,3}^{2\beta_\gamma} \Gamma_{grb,2}^4 t_{\gamma,1}^2} \;
        \frac{t}{1\ks} \,\epsilon_9^{2\beta_\gamma-1} \,.
\label{tausc}
\eeq

 In addition, the scattered photons can create pairs with GRB photons. The lab-frame 
energy of a GRB photon emitted at an angle $\theta \gt \Gamma_{grb}^{-1}$ relative to 
the radial direction of flow is a factor $(\Gamma_{grb} \theta)^2$ smaller than if the
photon were released at $\theta \lt \Gamma_{grb}^{-1}$. Then the condition for pair-formation 
leads to a minimum energy of the target photon (as measured by the observer if that photon 
were emitted at $\theta \lt \Gamma_{grb}^{-1}$) that is independent of $\theta$: 
\beq
 \epsilon_t^{(grb)} = 0.60\,[(z+1)/3]^{-2} \Gamma_{grb,2}^{-2} \epsilon_9^{-1}
               \; {\rm MeV} \;. 
\label{etgrb}
\eeq
 Further, it can be shown that, owing to the scattering outflow being behind the GRB 
source, the scattered GeV photons can interact only with burst photons that have been 
emitted at an angle larger than $\theta_{min}(t_{lag}) = (ct_{lag}/R_\gamma)^{1/2}$, 
with $ct_{lag}$ being the separation between the place of scattering and the GRB source 
and $R_\gamma = \Gamma_{grb}^2 ct$ the radius of the latter. 
For $\Gamma_{sc} \gt \Gamma_{grb}$, photons scattered by the fluid at $ct_{lag}$ behind
the GRB source arrive at observer at time $t=t_{lag}$, thus $\theta_{min}(t) \simeq 
(t/t_\gamma)^{1/2} \Gamma_{grb}^{-1} = 0.1\, (t_3/t_{\gamma,1})^{1/2} \Gamma_{grb,2}^{-1}$. 
Taking into account that the lab-frame distribution GRB photons with angle $\theta$ 
relative to the radial direction of motion 
is $dN_{grb}/d\Omega \propto [\Gamma_{grb}(1-B\cos\theta)]^{-2}$ (with $B$ the speed 
the GRB source), the photons emitted at $\theta \gt \theta_{min}$ are a fraction 
\beq
 f(t) = (\Gamma_{grb} \theta_{min})^{-2} = t_\gamma/t
\eeq
of the number of GRB photons
\beq
 N_{grb} = \frac{4\pi d_l^2(z)}{(z+1)^2} t_\gamma F_p
         = 10^{59.3} \z3^2 \phi_{-5} \epsilon_{p,5}^{-1} \;.
\label{Ngrb}
\eeq
Thus, the number of GRB photons above the threshold energy $\epsilon_t^{(grb)}$ that can 
reach the test photon is $N(>\epsilon_t^{(grb)}) = f (\epsilon_t^{(grb)}/\epsilon_p)^
{-\beta_\gamma} N_{grb}$. 
Then, for the scatterer radius given in equation (\ref{Rgrb}), equations (\ref{taugg}) 
and (\ref{etgrb})--(\ref{Ngrb}), lead to the following optical thickness to pair-formation 
on burst photons 
\beq
 \tau_{sc,grb} \siml 10^{1.5-0.8\beta_\gamma} 
   \h1 \z3^{\h1 2\beta_\gamma \hh +4} \h1 \frac{\phi_{-5}}{t_{\gamma,1}} 
    \frac{ \epsilon_{p,5}^{\beta_\gamma-1} } { \Gamma_{grb,2}^{2\beta_\gamma+4} } 
   \h1 \left(\hh\frac{t}{1\ks} \hh \right)^{\h1 -1} \h1 \epsilon_{-9}^{\beta_\gamma} \;.
\label{taugrb}
\eeq

 Equations (\ref{tausc}) and (\ref{taugrb}) show that GeV afterglow photons may be lost 
to pair-formation on upscattered photons or on burst emission. The energy above which 
this processes reduces the high-energy flux is strongly dependent on the Lorentz factor 
of the GRB source and, for pair-creation on the scattered emission, also on the Lorentz 
factor of the scattering outflow. 
 
 The KN effect reduces the GeV flux at photon energies approaching $\epsilon_* = 
\Gamma_{sc} \gamma_e m_e c^2 /(z+1) = 2 [(z+1)/3]^{-1}$ $(\Gamma_{sc}\gamma_e)_4$GeV. 
The KN cutoff energy $\epsilon_*$ is above the peak energy $\epsilon_p^{(sc)}$ of the 
scattered GRB emission (equation \ref{scnup}) if the scattering parameter $\Gamma_{sc} 
\gamma_e$ is less than $\Gamma_{grb}^2 (m_e c^2)$ $[(z+1)\epsilon_p]^{-1} = 2\times 10^4 
[(z+1)/3]^{-1}\Gamma_{grb,2}^2 \epsilon_{p,5}^{-1}$, for which $\epsilon_p^{(sc)} \lt 
3\, [(z+1)/3]^{-2}\Gamma_{grb,2}^2 \epsilon_{p,5}^{-1}$ GeV. 
The cut-off at $\epsilon_*$ is sharp only if the scattering electrons are cold 
($\gamma_e=1$); in this case, owing to that $\Gamma_{sc}$ is most likely below $10^4$,
the KN effect truncates the scattered spectrum above its peak energy.

\section{Conclusions}

 We have assessed the ability of four mechanisms related to GRBs and their afterglows
to produce LAT-detectable GeV emission during the early afterglow phase ($t \sim 1$ ks). 
We note that the expected GeV fluxes were calculated from afterglow and GRB observables 
(X-ray plateau flux, duration, and spectral slope; burst fluence, duration, and spectral 
properties), keeping to a minimum the use of specific model parameters that are not
well determined.

 Except for the hardest or longest \& brightest GRBs, the large-angle emission 
released during the burst and arriving later at observer should not be detectable to LAT. 
 With the exception of the hardest X-ray afterglows, detectable GeV emission is not 
expected from the same emission process that yields the X-ray emission, particularly if 
the X-ray afterglow is synchrotron emission from the forward shock. 

 The other two mechanisms discussed rely on the upscattering (either in the forward-shock
or an other part of the relativistic outflow) of lower energy photons into the GeV range
and offer better prospects for detection with LAT. 

 The self-Compton scattering of forward-shock synchrotron photons is expected 
to be above LAT's 1 ks sensitivity for X-ray afterglows with plateaus 
brighter than the average. This mechanism can be identified based on that the spectral 
slope of the GeV emission should be the same as for the X-ray afterglow ($\beta_{GeV} = 
\beta_x$). In addition the decay rates of the fluxes at these two energies should
satisfy $\alpha_{GeV} = \alpha_x+(\beta_x+1)/2$ (equation \ref{ic}). Pair-formation 
may reduce the self-Compton scattering flux at 1 GeV (equation \ref{tauic}) if the 
peak energy of the self-Compton emission spectrum is below 100 keV or of the X-ray
plateau emission is hard and extends well above 100 keV (equation \ref{tausy}).

 Bulk-scattering of the forward-shock emission by a delayed outflow can explain 
the X-ray plateaus observed in many Swift afterglows if the scattering outflow is 
pair-rich. The same mechanism may produce GeV emission that LAT can detect if the 
afterglow X-ray emission is brighter and harder than average. Scattering of the 
burst prompt emission may also produce detectable
GeV emission, particularly for bursts brighter than average and afterglows with 
short-lived X-ray plateaus (equation \ref{grbsc}). The upscattered emission should 
have the same spectral slope as the burst ($\beta_{GeV} = \beta_\gamma$) but can 
exhibit a variety of decays, depending on the radial distribution of mass and Lorentz 
factor in the scattering outflow. For this mechanism, pair-formation on either the 
burst photons or the upscattered emission itself reduces the GeV flux above a photon 
energy that is strongly dependent on the Lorentz factors of the GRB source and 
scattering outflow (equations \ref{tausc} and \ref{taugrb}). Consequently, the
detection of photons above 1 GeV during the early afterglow can be used to 
set lower limits on these Lorentz factors, similar to how it was done for the 
GRB source Lorentz factor using the detection of 0.1--10 GeV photons during
the burst by CGRO/EGRET (e.g. Fenimore, Epstein \& Ho 1993, Baring \& Harding 1997, 
Lithwick \& Sari 2001).

 For an effective area of 8000 ${\rm cm^2}$, LAT's sensitivity corresponds to several 
GeV photons detected in 1 ks. Thus, for most GeV afterglows, the high-energy light-curve 
will be too crude for a determination of its power-law decay rate. As the spectral 
slope can be determined with a modest number of collected photons, comparing the 
slope of the GeV spectrum with those of the burst and X-ray afterglow emission 
provides a better way to identify the mechanism that produced the high-energy afterglow. 
For GRB 940217, the only burst for which GeV afterglow emission was detected so far, 
EGRET observations (Hurley et al 1994) show a hard component at 0.1--4 GeV, with 
$F_\epsilon \propto \epsilon^{1/3}$ during the burst and, possibly, during the afterglow,
at 5 ks. The hardness of the GeV emission suggests that it is the burst emission below 
its spectral peak upscattered by a lagged outflow.

 We conclude by noting that, within the framework of the GeV afterglow emissions 
analyzed here, LAT will detect high energy photons under favourable conditions.
GRBs dimmer or harder than the average and X-ray afterglows with dim and long 
plateaus could be accompanied by GeV emission that is too faint to be detected by LAT.

\section*{Acknowledgments}
 The author acknowledges the support received from NASA Swift GI grant NNG06EN00I and 
LANL 20050161DR funding of Raptor

\end{document}